\documentclass[11pt]{article}
\usepackage[margin=1.5cm]{geometry}
\usepackage{titlepic}
\usepackage{graphicx}
\usepackage{float}
\usepackage{times}
\usepackage{url}
\usepackage{enumitem}
\usepackage[table]{xcolor}
\usepackage{csquotes}
\usepackage{multicol}
\usepackage{titlesec}
\usepackage[notes, backend=biber]{biblatex-chicago}
\bibliography{confidencebuilding}
\usepackage[bitstream-charter]{mathdesign}
\usepackage[T1]{fontenc}
\usepackage{setspace}
\usepackage{multirow}
\usepackage{array}
\usepackage{caption}
\usepackage{subcaption}
\usepackage{booktabs}
\usepackage{longtable}
\usepackage{changepage}
\usepackage{textcomp}
\usepackage{hyperref}

\usepackage{tocloft}

\renewcommand\thesection{\arabic{section}}

\usepackage{titlesec}
\titleformat{\section}
  {\huge}{\thesection}{2em}{}
\usepackage{titlesec}
\titleformat{\subsection}
  {\Large}{\thesubsection}{1em}{}

\renewcommand\thesection{\arabic{section}}
\makeatletter
\def\blfootnote{\xdef\@thefnmark{}\@footnotetext}
\makeatother

\begin{document}

\title{\Huge Confidence-Building Measures for Artificial Intelligence: Workshop Proceedings\blfootnote{All authors provided substantive contributions to the paper through sharing their ideas as participants in the workshop, writing the paper, and/or editorial feedback and direction. The first two authors are listed in order of contribution, and the remaining authors are listed alphabetically. Some workshop participants have chosen to remain anonymous. The claims in this paper do not represent the views of any author's organization. For questions about this paper, contact Sarah Shoker at sshoker@openai.com and Andrew Reddie at areddie@berkeley.edu.}}

\renewcommand*{\thefootnote}{\fnsymbol{footnote}}
\setcounter{footnote}{0}

\author{Sarah Shoker$^{1\footnote{Significant contribution, including writing, providing detailed input for the paper, research, workshop organization, and setting the direction of the paper.}}$, Andrew Reddie$^{2\setcounter{footnote}{6}\footnote{Significant contribution, including providing detailed input for the paper, research, workshop organization, and setting the direction of the paper.}}$, Sarah Barrington$^{2}$, Ruby Booth$^{3}$, Miles Brundage$^{1}$, Husanjot Chahal$^{1}$,\\ Michael Depp$^{4}$, Bill Drexel$^{4}$, Ritwik Gupta$^{2}$, Marina Favaro$^{5}$, Jake Hecla$^{2}$, Alan Hickey$^{1}$,\\ Margarita Konaev$^{6}$, Kirthi Kumar$^{2}$, Nathan Lambert$^{7}$, Andrew Lohn$^{6}$, Cullen O'Keefe$^{1}$, Nazneen Rajani$^{7}$,\\ Michael Sellitto$^{5}$, Robert Trager$^{8}$, Leah Walker$^{2}$, Alexa Wehsener$^{9}$, Jessica Young$^{10}$
\vspace{0.3in}\\
\small{$^1$OpenAI, $^2$University of California, Berkeley, $^3$Berkeley Risk and Security Lab,}\\\small{$^4$Center for a New American Security, $^5$Anthropic, $^6$Center for Security and Emerging Technology,}\\\small{$^7$Hugging Face, $^8$Centre for the Governance of AI, $^9$Institute for Security and Technology,}\\\small{$^{10}$Microsoft} }

\date{August 2023}

\maketitle

\begin{abstract}\noindent Foundation models could eventually introduce several pathways for undermining state security: accidents, inadvertent escalation, unintentional conflict, the proliferation of weapons, and the interference with human diplomacy are just a few on a long list. The Confidence-Building Measures for Artificial Intelligence workshop hosted by the Geopolitics Team at OpenAI and the Berkeley Risk and Security Lab at the University of California brought together a multistakeholder group to think through the tools and strategies to mitigate the potential risks introduced by foundation models to international security. Originating in the Cold War, confidence-building measures (CBMs) are actions that reduce hostility, prevent conflict escalation, and improve trust between parties. The flexibility of CBMs make them a key instrument for navigating the rapid changes in the foundation model landscape. Participants identified the following CBMs that directly apply to foundation models and which are further explained in this conference proceedings: 1. crisis hotlines 2. incident sharing 3. model, transparency, and system cards 4. content provenance and watermarks 5. collaborative red teaming and table-top exercises and 6. dataset and evaluation sharing. Because most foundation model developers are non-government entities, many CBMs will need to involve a wider stakeholder community. These measures can be implemented either by AI labs or by relevant government actors.

\end{abstract}

\vspace{1in}

\renewcommand*{\thefootnote}{\arabic{footnote}}
\setcounter{footnote}{0}

\newpage
\section{Introduction}

Foundation models could eventually introduce several opportunities for undermining state security: accidents, inadvertent escalation, unintentional conflict,\footnote{In this context, accidents occur when AI systems malfunction. Inadvertent escalation happens due to  inappropriate use of AI systems by leaders or operators that intensify situations. Unintentional conflict occurs when uncertainties in algorithm behavior hinder the ability of states to signal effectively to adversaries, potentially increasing the likelihood of conflict despite the ultimate intentions of involved states. \autocite{horowitzLeadingArtificialIntelligence2021}} the proliferation of weapons,\autocite{GPT4SystemCard2023} and the interference with  human diplomacy are just a few on a long list.\autocite{wardShaheenAdminGet2023} Meanwhile, new defense and security actors continue to develop foundation model capabilities,\autocite{PalantirArtificialIntelligence, DonovanAIpoweredDecisionmaking, milmoUKInvest900m2023, dingRecentTrendsChina2023} increasing the risk of an international crisis even further.

The \textit{Confidence-Building Measures for Artificial Intelligence} workshop hosted by the Geopolitics Team at OpenAI and the Berkeley Risk and Security Lab (BRSL) at the University of California brought together participants from AI labs, government, academia, and civil society to propose tools and strategies to mitigate the potential risks introduced by foundation models to international security. By foundation models, we mean models that use vast amounts of data, self-supervision and deep learning methods which “can be adapted…to a wide range of downstream tasks.”\autocite{bommasaniOpportunitiesRisksFoundation2022} The workshop included a mix of presentations and breakout groups, where participants had the opportunity to design possible confidence-building measures (CBMs). Together, participants identified the following CBMs that directly apply to foundation models:

\begin{itemize}
    \item crisis hotlines
    \item incident sharing
    \item model, transparency, and system cards
    \item content provenance and watermarks
    \item collaborative red teaming exercises
    \item table-top exercises
    \item dataset and evaluation sharing
\end{itemize}

Popularized during the Cold War, CBMs represent “measures that address, prevent, or resolve uncertainties among states. Designed to prevent the escalation of hostilities and build mutual trust among erstwhile adversaries, CBMs can be formal or informal, unilateral, bilateral, or multilateral, [such as] military or political, and can be state-to-state or non-governmental.”\autocite{ConfidenceBuildingMeasuresCrossStrait} Because states do not have perfect information about the capabilities or intentions of their allies and adversaries, formal and informal rules can establish predictability around state behavior, which in turn has the potential to reduce misunderstandings and miscommunications between state governments. This is in the interest of all parties.

Historical examples of CBMs include direct call lines between countries to communicate during nuclear crises, reporting on weapon transfers between states, inviting observers to witness military exercises that an outside nation might otherwise construe as threatening, establishing clear “rules of the road” for how adversarial navies should interact on the high seas in peacetime, data exchanges on troop movements such as those mandated by the Treaty on Conventional Forces in Europe, or on-site monitoring of technology capabilities. In contrast to domestic or regional AI regulations that govern the relationship between companies and consumers, CBMs target and address the risks associated with state-to-state interactions by introducing predictability into a typically opaque international environment. While CBMs can target the prevention of a range of harms, workshop participants focused on CBMs that mitigate human rights abuses, the proliferation of unconventional weapons, and escalation due to misperceptions exacerbated by foundation models.

Defense strategies now routinely address the risks and opportunities associated with artificial intelligence, with some governments and think tanks calling explicitly for confidence-building measures.\autocite{ChapterNSCAIFinal} Yet with the notable exception of the United Kingdom’s Integrated Review Refresh 2023, most governments have not fully grappled with the implications of military AI, much less foundation models.\footnote{Page 56 \autocite{sunakIntegratedReviewRefresh2023}} Though many existing defense documents do not directly target foundation models, governments can still fold the CBMs identified in these proceedings into existing AI commitments, such as the U.S Government’s Political Declaration on Responsible Military Use of Artificial Intelligence and Autonomy.\autocite{bureauofarmscontrolPoliticalDeclarationResponsible2023}

Building on the literature addressing the risks of AI to international security, this workshop focused on generating practical CBMs that apply to foundation models. The CBMs identified in these proceedings are not exhaustive or equally feasible in today’s international climate. Where appropriate, we outline political and technical limitations that could interfere with the CBM’s success.

\subsection{A Note on Terminology:  Foundation Models, Generative AI, and Large Language Models}

For the sake of brevity, we use the term ‘foundation model’ to refer to both base and fine-tuned models, generative AI, and large language models. Where appropriate, we identify the specific type of AI model the CBM is meant to address. The terms foundation model, large language model, and generative AI are often used interchangeably, but there are significant, if imprecise, differences between these terms. As Helen Toner notes, these terms do not have “crisp boundaries…[but]...have emerged as attempts to point to a cluster of research directions and AI systems that have become especially noteworthy in recent years.”\autocite{tonerWhatAreGenerative2023}

Foundation models are built using deep learning and self-supervision learning methods and use vast amounts of data which, according to a 2022 paper by Rishi Bommasani et al. at Stanford University,  “can be adapted (e.g. fine-tuned) to a wide range of downstream tasks.”\autocite{bommasaniOpportunitiesRisksFoundation2022} The large amount of data and computational power used to train foundation models have led to impressive improvements across a variety of domains.\footnote{\textit{ibid.}}

While foundation models are often associated with generative AI applications like language and imagery (see below), these models can also be applied to domains such as robotics, human-machine interaction, reasoning, and sentiment analysis. On the other hand, generative AI is a narrower category of AI that includes models and algorithms capable of generating media. These models produce content like text, audio, imagery, and software code. Many public-facing models that are available today have already been fine-tuned on a foundation model. For example, ChatGPT models are fine-tuned on foundation models called GPT-3.5 and GPT-4, while Stability AI uses foundation models like StableLM to generate imagery.

\subsection{Why do we need confidence-building measures for foundation models?}

There is no shortage of historical crises where misperception or miscommunication led to military escalation that neither side wanted.\footnote{Misperception continues to be a popular research area for scholars of military conflict, and some researchers suggest that the academic existence of international relations is fundamentally linked to managing problems related to information asymmetry and the anarchical conditions that make misperception possible.} Misperception plays a prominent causal role in the bloodiest wars of the 20th century, whether that be in both World Wars, Cold War ‘proxy’ conflicts like Vietnam, Korea, and Afghanistan, or more recent 21st century conflicts like the Second Gulf War and Syrian Civil War. There are  ample cases of militaries mistakenly targeting civilian planes and killing most or all civilians onboard,\autocite{depasqualeCivilianPlanesShot2020} and there are numerous historical false positives that only narrowly avoided nuclear exchange.\footnote{For a full list of nuclear false alarms, please visit compendium of events.\autocite{CloseCallsNuclear2015}}

The flexibility of CBMs make them a key instrument for navigating the rapid changes in the foundation model landscape. AI is a general purpose “enabling technology” rather than a military technology in and of itself.\footnote{Page 6 \autocite{puscasConfidenceBuildingMeasuresArtificial2022}} For instance, current rule-making at the United Nations Convention on Certain Conventional Weapons (UN CCW) focuses on weapons identified by the forum,\autocite{ConventionCertainConventional} which excludes many AI applications–such as generative AI–that are not obviously categorized as a ‘weapon’ but that can nevertheless influence the direction of international conflict. In particular, their non-binding, build-as-you-go nature allows the CBMs to grow in specificity as the technology necessarily evolves. This is essential, since it is not obvious what capabilities foundation models possess after they are trained and new capabilities are often revealed only after further red teaming and conducting safety evaluations. Though several benchmarks exist for assessing foundation models, they overwhelmingly point to rapid improvement in domain knowledge and deduction.\autocite{duaDROPReadingComprehension2019, hendrycksMeasuringMassiveMultitask2023, PapersCodeMMLU} These capabilities are already associated with international security risks like providing information on the construction of conventional and unconventional weapons.\footnote{Page 12 \autocite{GPT4SystemCard2023}}

CBMs do not overrule or subvert important efforts at fora like the United Nations and can act as an accompaniment to ongoing international regulatory discussions. CBMs are, however, uniquely  equipped to target risks associated with foundation models due to the speed of their innovation and proliferation. In comparison to formal rules or international treaties, CBMs can lower coordination costs (such as time and money spent on bargaining) by reducing the number of negotiating parties involved in discussions. CBMs are often voluntary, which can incentivize participation from parties who are reluctant to risk the full weight of their national credibility on formal treaties. CBMs are more easily modified (and discarded).\autocite{horowitzFutureMilitaryApplications2020} CBMs can also ‘start small’ and build into formal rules, an especially useful feature in a low-trust international environment.\autocite{horowitzLeadingArtificialIntelligence2021}

Model performance and model safety are also separate research pursuits, meaning that the performance of foundation models can improve with little change to their safety profile. A large language model that can generate information about nuclear physics is an example of a capability, while a large language model that refuses a user request to output specific details about bomb-building is an example of a safety mitigation. To date, AI labs have tackled the gap between model performance and safety by investing in a range of sociotechnical measures. Such measures include research into interpretability and alignment,\autocite{wuRecursivelySummarizingBooks2021, billsLanguageModelsCan2023, baiConstitutionalAIHarmlessness2022a, elhageToyModelsSuperposition2022, shahGoalMisgeneralizationWhy2022, zhouLeasttoMostPromptingEnables2023, zieglerAdversarialTrainingHighstakes2022, cohenAdvancedArtificialAgents2022} public disclosure of risks through system cards\autocite{GPT4SystemCard2023} and transparency notes,\autocite{chrishmsftTransparencyNoteAzure2023} delaying the release of models until sufficient safety mitigations have been implemented,\footnote{Page 19 \autocite{GPT4SystemCard2023} and \autocite{MicrosoftTuringAcademic}} and open-sourcing evaluations\autocite{OverviewC2PA, englandAMPAuthenticationMedia2020, Evals2023} and provenance research.\footnote{For example, as part of its Content Authenticity Initiative (CAI), Adobe open-sourced JavaScript SDK and Rust SDK, which is “designed to let developers build functions displaying content credentials in browsers, or make custom desktop and mobile apps that can create, verify and display content credentials.”\autocite{gowranAdobeLaunchesOpensource2022}} Despite these efforts, the machine learning community is in general consensus that harm mitigations need further improvement to keep up with the rapidly increasing performance of LLMs.\autocite{CoreViewsAI, leikeOurApproachAlignment2022, glaeseImprovingAlignmentDialogue2022a}

This landscape is further challenged by typical state behavior at the international level. States are often reluctant to engage in cooperative security agreements that require too much transparency into national capabilities. They are even less likely to place limits on the development of their own capabilities in the absence of any guarantees that their adversaries will do the same.\footnote{For example, France’s Defence Ethics Committee advised the continuation of research into autonomy and weapons systems, citing, among other reasons, the need to  “counter enemy development of LAWS; and…to be able to defend ourselves against this type of weapon in the likely event of their use by an enemy State or terrorist group against our troops or population.”\autocite{OpinionIntegrationAutonomy2021}} However, because performance and safety research are two different research streams, it is possible to coordinate on security while limiting availability of research into performance improvements. This unintended silver-lining is known to AI labs, which is why commercial labs are often willing to open-source safety research into evaluations and provenance technologies.

\subsection{An Overview of CBMs for Foundation Models}

Drawing from the list published by the United Nations Office of Disarmament Affairs, these proceedings organize CBMs under four categories: communication and coordination, observation and verification, cooperation and integration, and transparency.\autocite{RepositoryMilitaryConfidencebuilding} These categories are not discrete; many CBMs can comfortably fit into more than one category.

Because most foundation model developers are non-government entities, many CBMs will need to involve a wider stakeholder community. These measures can be implemented either by AI labs or by relevant government actors.\footnote{We opted to exclude a discussion on cyber risks from the scope of this paper since legal advances published in the Tallinn Manuals, NATO announcements on what counts as ‘cyber war,’ and norm setting at the UN Group of Governmental Experts on state behavior in cyberspace means that the topic deserves its own devoted forum.} Throughout the paper, we provide examples of adjacent technologies that have contributed to international crises, with the understanding that these examples can help us better anticipate the risks posed by foundation models, which are currently  nascent or empirically unconfirmed.

\subsection{Communication and Coordination}

Communication and coordination CBMs reduce misunderstandings and misperceptions that, if left unaddressed, could escalate into conflict. The workshop identified two communication and coordination challenges that could be remedied using communication and coordination CBMs: misperceptions about authenticity of the content, and misperceptions concerning who authorized a decision.

First, on the topic of content authenticity, several workshop participants reiterated that foundation models and, specifically, generative AI, can be used to perpetuate ‘truth decay,’ or increased public distrust towards the information reported by political leaders and other experts. That distrust, in turn, complicates reporting on international events and crises.\autocite{goldsteinGenerativeLanguageModels2023a, oltermannEuropeanPoliticiansDuped2022} For example, in March 2022, a widely circulated deepfake video on social media showed Ukrainian President Volodymyr Zelenskyy instructing soldiers to surrender to Russian forces.\autocite{allynDeepfakeVideoZelenskyy2022} Individuals may soon speak with \textit{interactive} deepfakes, where the deepfake is both able to pause appropriately for the other speaker and use predictive modeling and synthetic audio to carry on a conversation.\autocite{horvitzHorizonInteractiveCompositional2022a} And we could see the use of compositional deepfakes–not just one fake video or image, but many of them–released over time in between real events, to create a synthetic history that seems believable.\footnote{\textit{ibid.}}

Second, strong communication and coordination CBMs allow human actors to account for the ambiguity an AI injects into a system or team.\autocite{horowitzLeadingArtificialIntelligence2021} AI systems are often designed with the intention of  supporting or augmenting human decision-making, making it challenging to disentangle the contributions of human operators. AI systems may also generate outputs or decisions that can be misinterpreted or misunderstood by human operators or other AI systems; in some cases the integration of AI in human-machine teams\footnote{We define ‘human machine’ team as “a relationship–one made up of at least three equally important elements: the human, the machine, and the interactions and interdependencies between them.”\autocite{BuildingTrustHumanmachine}} can obfuscate whether AI was the (inadvertent, accidental, or intentional) cause of military escalation. A case in point is the 1988 tragedy of Iran Air Flight 655, which was targeted by an Aegis cruiser–the most sophisticated anti-aircraft weapon system at the time–on the order of the USS Vincennes, killing 290 civilians. The accident was blamed on a number of factors: the Aegis incorrectly identified the commercial airliner as a military aircraft; the commander of the Vincennes was characterized as being unnecessarily aggressive in a high-pressure atmosphere prone to misinterpretation; a nearby US navy ship, the USS Sides, had a human commander who correctly deduced that Iran Air Flight 655 was a civilian aircraft, but believed the Aegis's identification system to be technologically superior to his own human judgement and did not share his assessment with the Vincennes.\autocite{H0201USSVincennes, FormalInvestigationCircumstances1988} The Aegis radar system did eventually identify the Iran Air Flight 655 as a civilian aircraft, but the human operators chose to accept the first reading. The Iran Air Flight 655 accident features many of the challenges that exist in today's human-machine teams: overtrust and a reluctance to contest the decisions made  by the system, misunderstanding the threat due to existing geopolitical hostilities, and cherry-picking evidence to support one's interpretation of events. The introduction of AI to this atmosphere, which promises to increase the speed of targeting and analysis using a black-boxed technology, makes it even more necessary to identify communication pathways to prevent accidents.

\textit{Hotlines}

The ability to interpret human intentions can become more challenging when communication integrates with or is supplanted by a non-human entity. Hotlines can assist with clarifying the ‘who’ or ‘what’ was responsible for military escalation, and for clarifying red lines\autocite{wolfBackingWhyRed2016} to avoid crossing them in the first place.\footnote{For more information on how hotlines can clarify red lines, see: \autocite{whitakerWhenRussianHackers2018}} Workshop participants noted that competitor states could establish communication links to reduce friction during political crises, building on state-to-state hotlines that exist today for the management of military crises.

Despite their prominent role in mitigating nuclear crises, recent political events have underscored the reality that security norms will inform when parties use–or refuse–a phone call. This point was made especially evident during the February 2023 crisis involving a Chinese spy balloon traveling across the United States, and the subsequent refusal by the People’s Liberation Army (PLA) to answer a hotline call from U.S Defense Secretary Lloyd Austin. Immediately following the crisis, researchers offered several explanations for the PLA’s behavior that pointed to a discrepancy between how both military powers interpreted the threat landscape. Some stated that the PLA viewed CBMs and transparency as “disadvantageous” and a normalization “of increasingly brazen behavior.” Another researcher stated that U.S military norms prize the avoidance of military escalation, while “[i]n the Chinese system, the impulse is to not be blamed for a mistake” or to be the person who reports the message to their political or military leaders.\autocite{lafranchiUSChinaConundrumCan2023} It is worth noting that hotline usage becomes even more complicated in a world with three major military powers, where incentives could exist for one actor to exploit crisis communication between the two other states.

The successful use of hotlines may require that parties share common values about the risks of foundation models and a mutual belief that CBMs reduce the risk of unnecessary military escalations. States routinely disagree about the severity of threats and will pursue technologies to keep their own borders safe, even at the expense of global security. Other CBMs listed in this document, such as collaborative red teaming, emergency-response tabletop games, and incident sharing, can supply the necessary data for assessing the risk landscape while reinforcing the idea that CBMs will not undermine any single state’s security. As an area of future study, participants recommended research on understanding policymaker perceptions about foundation model risks to international security and ensuring that incentives for participating in CBMs address country-specific social values.

\textit{Incident Sharing}

Incident-sharing is a common practice across sectors where public safety is paramount, such as electric vehicles, cybersecurity, aviation, and healthcare. Information sharing about security incidents or ‘near misses’ is used to improve safety and reduce the likelihood of new accidents. With regards to autonomy in military systems, Michael Horowitz and Paul Scharre have previously suggested that an “‘international autonomous incidents agreement’ that focuses on military applications of autonomous systems, especially in the air and maritime environments…would reduce risks from accidental escalation by autonomous systems, as well as reduce ambiguity about the extent of human intention behind the behavior of autonomous systems”.\autocite{horowitzAIInternationalStability2021} This problem is documented in technology modernization for defense. For example, the introduction of touchscreen controls to the USS John S McCain, combined with crew confusion about the different settings associated with these controls, contributed to the largest maritime accident involving the US Navy in the last 40 years and left 10 sailors dead.\autocite{NTSBAccidentReport2019}

Open-source AI incident-sharing initiatives already exist, such as the AI, Algorithmic, and Automation Incidents and Controversies (AIAAIC) and AI Incident Databases.\autocite{AIAAIC, AIIncidentsDatabase}  As of April 2023, these open-source databases primarily featured journalistic investigations, which sometimes include incidents on generative AI and international security, like the recent deepfake of the US Pentagon explosion.\autocite{Incident543Deepfake2020} Participants suggested a comparable database for international security incidents caused by foundation models, with a possible focus on unusual vulnerabilities and emergent model behaviors.

Workshop participants raised several questions that remain unresolved. Namely, it is unclear which model behaviors and misuses would qualify as an “incident,” the incentives for parties to participate in an incident-sharing agreement, and how those parties can assure accurate reporting while respecting intellectual property rights and user privacy. A distinction might exist between new and dangerous model capabilities versus the large-scale misuse of the model. The former category could include behaviors linked to improvements of the model such as the ability to manipulate users or design a synthetic biochemical agent. The latter category could entail large-scale misuse campaigns, such as using models to create spam or disinformation.

Other industries resolve these challenges through data anonymization and encryption, trusted third-party agreements, access controls, NDAs, and security audits. Some types of incident sharing can leverage existing professional relationships between labs and could be as simple as hosting an informal meeting amongst parties. However, incident-sharing may require a multilateral entity that coordinates incident collection across multiple parties. Workshop participants noted that an AI incident-response entity could be analogous to existing Computer Emergency Response Teams (CERT) found in the international cybersecurity domain.\footnote{For examples on national CERTs, please see: \autocite{USCERTUnitedStates, CERTEUComputerEmergency}}

\section{Transparency}

AI systems may produce unintended outcomes due to biases in training data, algorithmic errors, or unforeseen interactions with other systems. To name a few examples: foundation models used to summarize ISR data can introduce artifacts into the data that impacts a military response. Plausible outputs that are actually false, known as "hallucinations",\autocite{tamGentleIntroductionHallucinations2023} can be difficult to detect in a fast-paced and high-pressure military environment. Moreover, labeling practices can contribute to bias by privileging some worldviews over others, a serious risk for intelligence analysts conducting even routine tasks like report retrieval and summarization.\footnote{Scholars working within science and technology studies frequently note that the labeling of datasets reveal the political preferences and biases of the labellers, which has direct consequences for the model’s performance. As Kate Crawford and Trevor Paglen note, “the automated interpretation of images is an inherently social and political project, rather than a purely technical one”\autocite{crawfordExcavatingAI}} Compounding this problem is that models do not perform equally well across languages and it is seldom clear whose values should be reflected in model generations. Finally, prompt injection attacks, a type of data poisoning and security exploitation, can alter model outputs.\autocite{ExploringPromptInjection2022} Prompt injection attacks are made easier when the adversary has access to the training data or model weights.

Some workshop participants also cited the problem of information overload. Even if accurate, too much information creates its own set of risks. States are often hesitant to escalate military activity because they are distrustful of their own predictions and intelligence about adversaries.\autocite{snyderDeterrenceDefense1961} For example, machine learning could improve sensors to a point which renders the sea domain overly transparent and erodes the deterrent capacity of second strike forces.\autocite{johnsonArtificialIntelligenceDrone2020a} In general, however, access to accurate information trends towards international stability.\autocite{jervisCooperationSecurityDilemma1978, fearonRationalistExplanationsWar1995, duelferChronicMisperceptionInternational2011} To address the challenges posited above, workshop participants explored a variety of confidence-building measures. These are outlined, in brief, below.

\subsection{Transparency Reports, Model and System Cards}

System cards are documents that detail intended use cases, limitations, and the results of red teaming,\footnote{Similarly, the \hyperlink{https://datanutrition.org/}{Data Nutrition Project} draws inspiration from nutrition labels found on food packaging. For an example from OpenAI, see \autocite{DallE2SystemCard}} comparable to documentation practices found in industries such as aerospace, medicine, and pharmaceuticals.\autocite{rajiClosingAIAccountability2020, mitchellModelCardsModel2019a} In domestic settings, proponents of system cards argue that they can help policymakers better understand the capabilities and limitations of AI systems, informing oversight and regulation.\autocite{mitchellModelCardsModel2019a} System cards do not require a third party to have access to the model itself, meaning that they can introduce transparency about capabilities while not revealing research details that would enable reverse-engineering of the model.

For foundation models used in defense domains, system cards should also include risks associated with human-machine interaction and overreliance, which can help outside observers interpret a system’s behavior in the event of an accident or escalation. (It is not always possible to know who or what is responsible for a system’s strange behavior.) For example, a 2021 UN Security Council report described a Turkish-made Kargu-2 drone in Libya as an instance where a lethal autonomous weapons system was deployed in violent conflict, a description that generated significant controversy in the international security community and highlighted the uncertainty involved with understanding the behavior of human-machine teams.\autocite{knightAutonomousWeaponsAre}

For best effect, system cards should be readable and easily accessible. Many of today’s system cards are found on code repository websites like Github, sites which tend not to be frequented by policymakers, and written in formats that those outside the field of machine learning can sometimes find inaccessible.

Like other measures found in these proceedings, there are limitations to model and system cards. Specifically, outside parties can experience difficulty verifying the results of model and system cards. Limitations can exist in non-adversarial and non-military contexts, too. Foundation models are often unavailable to third parties, with the exception of base models made available by companies that open source their models. When foundation models are made publicly available, they are often  refined through reinforcement learning from human feedback (RLHF), as seen in the InstructGPT models that power ChatGPT.\autocite{loweAligningLanguageModels2022} In general, publicly available fine-tuned models tend to be safer due to safety measures implemented after the base model has completed training.\footnote{Fine-tuning can also be used to make the model less safe, as actors could fine-tune base models on harmful or otherwise undesirable information. A recent example is the controversial release of ‘GPT-4Chan,’ where an open source model developed by Eleuther AI called GPT-J was finetuned on 4Chan forum posts. The model was hosted on the HuggingFace hub and even contained a model card before HuggingFace made the decision to restrict access and eventually remove the model. See: \url{https://huggingface.co/ykilcher/gpt-4chan/discussions/1}} Transparency reports, model and system cards often document the capabilities and limitations of the base model, making it difficult for third parties to replicate or otherwise validate the findings in these documents. This limitation is especially problematic in the context of international security, where adversaries may have several reasons to exaggerate or under-report the capabilities of their base models.

For this reason, model and system cards should be supported by other coordination activities, such as collaborative red teaming (explained in the section on ‘Cooperation, Collaboration, and Integration.’)

\subsection{Observation and Verification}

Parties can also agree to using empirical methods to observe and verify that actors are complying with agreements. These techniques do not normally guarantee full transparency, since states are reluctant to reveal the full scope of their military capabilities, as explained above. The Biological Weapons Convention (BWC), Chemical Weapons Convention (CWC), and the Treaty on the Non-Proliferation of Nuclear Weapons (NPT) all include mechanisms for third-party verification. Verification is also a key feature in agreements outside the UN. For example, the Open Skies Treaty allows signatories to fly observation aircraft to collect data on some military capabilities and activities.\autocite{TreatyOpenSkies1992}

The success of verification is often dependent on the availability of detection and monitoring technologies. For this reason, detecting AI misuse in the international context has emphasized restrictions on hardware, since software can easily proliferate and evade monitoring. U.S. efforts have so far focused on constraining the supply of semiconductors and semiconductor manufacturing materials through export controls (though export controls are not necessarily confidence-building measures.)\footnote{Laws that restrict the transfer of goods, technologies, and services to entities outside the country. Some export controls restrict this transfer to specific actors in another country (‘end-user control’) or restrict how the technology can be used (‘end-use’).} These controls have mostly targeted two countries: Russia due to its invasion of Ukraine, and the PRC, which the 2022 U.S. National Security Strategy identifies as the “only competitor with the intent, and, increasingly, the capacity to reshape the international order.”\footnote{Page 8 \autocite{biden2022USNational2022}} However, algorithmic improvements, fine-tuning, the wide availability of consumer-grade LLM APIs, and open source alternatives mean that hardware controls are likely to be insufficient for preventing misuse. Moreover, while technology denial can constrain the range of choices available to any particular state, it does not address the problem of states’ behavior with technologies already at their disposal.

\subsection{Content Provenance and Watermarking}

Content provenance and watermarking methods assist with the disclosure and detection of AI content and can reduce misperceptions by establishing norms around the use of generated content during international crises. Provenance and watermarking methods can improve traceability, alleviate concerns about the origin of the AI generated or edited content, and promote trust among parties. If properly vetted against adversarial manipulation, they can also help states use AI-generated products more confidently, knowing that the outcomes can be traced back to their source. States can verify if the AI systems deployed by other parties adhere to agreed-upon guidelines or restrictions, making it easier to address any potential violations. 

Content provenance is an ongoing and politically salient area of research, development, and adoption. For example, the Coalition for Content Provenance and Authenticity (C2PA), whose members include Adobe, Microsoft, Intel, BBC, Sony, and Truepic, is an industry-led initiative that develops technical standards for establishing the source and history of media content. Born to address the “prevalence of misleading information online,” the coalition also offers a set of technical specifications for developers and guidelines to help users reason through the provenance of media content. As per the C2PA specifications, provenance methods can be split between "hard" and "soft" bindings, with the former including methods for applying unique identifiers to data assets and other cryptographic methods.\footnote{See section 2.4: \autocite{C2PASecurityConsiderations}} For instance, a C2PA manifest using cryptographically-bound provenance can include information about the origin of a piece of content (such as the AI model and version used to create it) and edits made to the content over time. A full survey of AI provenance methods is out of scope for this paper, but is well worth further research to determine which methods can be applied to improve state-to-state interactions and how these efforts complement each other.\footnote{For a list of disclosure methods, please see \autocite{PAIResponsiblePractices}}

One of the most popular and readily-available AI disclosure methods in use today is "watermarking" (which the C2PA describes as a "soft" binding because they are more easily undermined in comparison to a "hard" binding.) Watermarking involves embedding low probability sequences of tokens into the outputs produced by AI systems, which can serve as a verification mechanism to confirm the authenticity and integrity of AI generations. Watermarks are traceable, meaning that they can enable parties to trace AI-generated outcomes back to their source system, thereby allowing stakeholders to identify which AI model was used and who was responsible for deploying it. However, watermarks are also accompanied by a severe restriction: they are not tamper-proof. For example, bad actors can use "paraphrasing attacks" to remove text watermarks, spoofing to infer hidden watermark signatures, or even add watermarks to authentic content.\autocite{sadasivanCanAIGeneratedText2023}

Because watermarking for large language models is a nascent area of research, the technique is mostly useful for photorealistic imagery, though watermarking for AI imagery also faces many limitations. Many AI-image generators that are publicly available today are already accompanied by watermarking methods, but these methods can be adversarially circumvented using post-processing methods on the image. For example, some watermarks for AI images can be removed through JPEG compression.\autocite{jiangEvadingWatermarkBased2023} Due to such limitations, provenance tools should be frequently red teamed to verify their resilience against adversarial tampering, and C2PA provides security guidance to protect against attackers trying to tamper with provenance methods.

Watermarking is publicly popular, as demonstrated by a Vox Media survey that found that 78 percent of American adults believe that AI-generated media should be clearly disclosed.\autocite{wongWhatAmericansAre2023} Provenance and watermarks for imagery and audio feature prominently in the recent commitments made by AI companies to White House.\autocite{EnsuringSafeSecure} Despite its popularity with the U.S. public, the adoption of disclosure methods could also be contentious for both commercial and political reasons. First, small developers may not have the resources to invest in and apply provenance and watermarking technology. For this reason, open provenance standards and open sourcing AI detection technologies should be encouraged to help reduce the cost of security. Second, AI developers may also be reluctant to use watermarks on imagery for fear of alienating their consumers. Third, and as seen in the cybersecurity domain, states prefer certain technologies precisely because the malicious behavior is difficult to trace back to the belligerent party. States often exploit technical and political ambiguities about "what counts" as an escalatory military behavior so that they can continue to engage in conflict below the threshold of war and avoid attribution.\autocite{ridAttributingCyberAttacks2015} Parties could exploit generative AI for the same reason, since it is currently unclear how the use of such models are interpreted by competitor states. While the proliferation of foundation models means that provenance and watermarking is unlikely to be applied evenly by all developers, states can commit–even unilaterally–to using such technologies in diplomatic and security activities.

\subsection{Policies and Procedures}

Rather than providing information about the models and systems that might be used, states could share information about the processes or procedures for assuring that they are safe. This could involve sharing baseline testing and best practices used to verify and validate AI-enabled systems. Since safety assurance is typically a mutually-aligned goal, some even envision that development of baseline testing techniques and procedures could become a collaborative effort among allies and adversaries.\autocite{wehsenerAINC3IntegrationAdversarial, morganMilitaryApplicationsArtificial2020}

In addition to testing, publishing the policies and procedures for acquiring and approving AI-enabled systems can also provide confidence that systems are developed responsibly without divulging intellectual property. This could involve disclosing the minimum standards for performance such as the safety-integrity levels that exist for other safety-critical systems.\autocite{lohnEstimatingBrittlenessAI2020} An option that reveals even less potentially sensitive information is to publicly name the responsible parties for approving the development, acquisition, and use of potentially worrisome capabilities. Even simply defining the capabilities that would require those extra approvals could help provide some clarity. DoD Directive 3000.09 does not satisfy all advocates, but it does make progress at providing clarity around some of these issues.

\section{Cooperation, Collaboration, and Integration}

Of course, many of the measures discussed above require AI labs and governments to collaborate and address the most proximate risks. Parties can coordinate security activities for the purpose of building trust and learning from one another. In higher trust environments, these activities can encourage transparency around military capabilities. In low trust environments, even simulation exercises can be difficult to organize.

\subsection{Collaborative Red Teaming Exercises}

Workshop participants advocated for collaborative red teaming, in (coincidental) alignment with the Biden Administration’s recent announcement on responsible AI innovation, which featured a “public evaluation of generative AI systems.”\footnote{In May 2023, The Biden Administration announced that major AI developers, including Anthropic, Google, Hugging https://www.overleaf.com/project/64b6a3eadcc06995e5dc3666Face, Microsoft, NVIDIA, OpenAI, and Stability AI, will participate in a public evaluation of generative AI systems at DEFCON 31's AI Village, with the aim of determining whether these models align with the Biden Administration’s “Blueprint for an AI Bill of Rights” and “AI Risk Management Framework.” See: \autocite{houseFACTSHEETBidenHarris2023}} Collaborative red-teaming in the United States is currently in development as a public transparency and multistakeholder activity being held at DefCon 2023. The event includes participation by several companies, including Google, Hugging Face, Microsoft, NVIDIA, OpenAI, and Stability AI. Multilateral exercises similarly exist for cybersecurity, such as the annual Locked Shields exercise hosted by NATO’s Cyber Defence Centre of Excellence (CCDCOE); in 2022, the red-team, blue-team exercise was attended by over two thousand cyber experts from thirty-two countries.\autocite{LockedShields} Unlike vulnerability discovery for cyber systems, red teaming foundation models often refers to capability discovery and require little background in machine learning. In turn, red team activities can improve emergency preparedness by exposing relevant stakeholders to the risks associated with foundation models.

\subsection{Table-Top Exercises}

Table-top exercises bring together stakeholders to simulate and discuss their responses to potential accidents or seemingly intractable solutions, improving crisis readiness and joint planning, and reducing the likelihood of misunderstandings during real-world conflicts. TTXs can also enhance coordination; states can develop a better understanding of each other’s emergency procedures, identify areas where their response mechanisms or capabilities need improvement, and share best practices. TTXs between competitors, such as in track 2 diplomatic contexts, can improve the mutual understanding of intentions and surface risks or scenarios that might not have been considered.

Like red-teaming exercises, international fora can act as hosts for these events. The CCDCOE could integrate foundation models into Locked Shields to explore the prospect of cyber vulnerabilities, while the United Nations Institute for Disarmament Research (UNIDIR) could conduct a red-teaming exercise linked to the aims of the CCW discussions on autonomy in weapons systems. Because table-top exercises often serve as pedagogical tools, they can also be thought of as a ‘training and education’ CBM.

\subsection{Dataset and Evaluation Sharing}

Dataset sharing allows for the integration of safety standards across labs. Not to be confused with incident sharing, AI labs can collaborate on "refusals" by sharing datasets that focus on identifying and addressing safety or ethical concerns in AI-generated outputs. In the context of foundation models, "refusals" refer to instances where the AI system intentionally does not generate an output or refrains from providing a response to a user query due to safety or ethical concerns. This may occur when the requested output could potentially lead to harmful consequences, promote misinformation, or violate the ethical guidelines and policies set by the AI developers. Sharing such datasets can contribute to the development of more robust, aligned, and responsible AI systems.

In the area of international security, these datasets can contain information related to dual-use scientific information and that are legal to share across parties, such as in domains like chemical, biological, radiological, and nuclear science (CBRN). Red teaming has demonstrated that the interaction between LLMs and CBRN can introduce new proliferation pathways, potentially empowering non-state threat actors.\footnote{See page 12, 16: \autocite{GPT4SystemCard2023}} Sharing datasets allows AI labs to establish and integrate common benchmarks for evaluating the effectiveness of refusal mechanisms in LLMs. Datasets and evaluation sharing can also help to improve red teaming in small labs that do not have the resources to contract a large group of external experts.

There are limitations to dataset and evaluation sharing, especially as they relate to issues like national security. These limitations involve the regulation of dual-use items, the technical shortcomings of refusals, and expected future improvements in foundation models. Some scientific information is regulated to prevent the spread of information that can be used for weapons proliferation. For example, the U.S. International Traffic in Arms Regulation (ITAR) establishes controls on the import and export of technologies on the United States Munitions List (USML). The USML includes some dual-use technologies, and the regulations include rules surrounding the distribution of information about those technologies to foreign actors,\autocite{InternationalTrafficArms2020} adding a layer of complexity to red teaming and the development of safety mitigations. As a result, many labs avoid this problem by red teaming information that is not controlled.

On a more general note, it is not yet clear if models will "discover" new biochemical compounds in the future and, if so, whether these discoveries may introduce new security vulnerabilities. Refusals target capabilities that have already been discovered, meaning they are a powerful but limited solution in the domain of biochemical discovery.

Despite these limitations, there are still benefits to sharing datasets that contain public, but sometimes difficult to find, information. Because information contained in these datasets are public, trialing dataset sharing today versus later is a comparatively lower risk endeavor, the main obstacle being the leveraging of scientific talent for capability discovery. In comparison, future data-sharing may be a higher risk activity if foundational models take on a greater role in biochemical discovery. And like all CBMs, the sharing of datasets demonstrates a commitment to transparency and responsible AI development, which can contribute to building trust among AI labs across countries, policymakers, and the broader public.

\section{Conclusion}

States are often reluctant to limit their technical capabilities. This is especially true in the face of heightened international competition and when confronting uncertainties around a new technology.\footnote{For example, France’s Defence Ethics Committee advised the continuation of research into autonomy and weapons systems, citing, among other reasons, the need to  “counter enemy development of LAWS; and…to be able to defend ourselves against this type of weapon in the likely event of their use by an enemy State or terrorist group against our troops or population.” \autocite{OpinionIntegrationAutonomy2021}} However, military interest in AI, and increasingly in foundation models and generative AI capabilities, has intensified the urgency of establishing an international code of conduct for state behavior, as the Political Declaration on Military Uses of Artificial Intelligence and Autonomy illustrates.\autocite{horowitzLeadingArtificialIntelligence2021} As Rebecca Hersman notes in her work on the potentially new escalation dynamics caused by emerging technologies, “[u]nlike traditional concepts of escalation, which suggest linear and somewhat predictable patterns from low-level crisis to all-out nuclear war, escalatory pathways in this new era of strategic competition will be less predictable.”\autocite{hersmanWormholeEscalationNew2020} While CBMs are not a complete solution for the international system’s various insecurities, they do offer a set of options for reducing the likelihood of violent conflict caused by misinterpretation and miscalculation.

Though this workshop was primarily designed for thinking about solutions, participants identified a number of risks that could undermine the feasibility of CBM adoption. First, workshop attendees highlighted the information disparity between technologists and policymakers. The speed of politics–and diplomacy, in particular–often lags behind the speed of capability development, compounding the challenges of establishing appropriate CBMs in step with emerging tools. Policymakers may struggle to negotiate or provide assurances to their counterparts in other countries if they are unaware of the capabilities that exist within their borders.

Participants called for an increase in candid multistakeholder conversations to alleviate this problem. While the CBMs for AI workshop served as a space to discuss the impact of foundation models on international security, multi-stakeholder opportunities have predominantly been sporadic and reliant on the initiative of voluntary contributors to organize what are often time-consuming meetings.

Second, there are different options with respect to who would coordinate implementation and adoption of these CBMs, each with tradeoffs and drawbacks. For example, incident sharing demands not just funding, but also a reliable third party with sufficient staff to manage intake and ensure database quality. Participants suggested a variety of mechanisms to address this coordination issue, ranging from the addition of new offices within existing government agencies like the U.S. Office of Science and Technology Policy and parallel agencies in other countries, to forming international institutions that would oversee compliance and distribute benefits for participating states and labs. Two sandbox groups independently noted the oft-used and potentially problematic analogy of an ‘IAEA for AI’ as a comparable entity. Workshop participants suggested that states that abide by monitoring and verification norms could gain access to data pools, with initial access granted to testing and evaluation (T\&E) infrastructure, followed by access to data and subsidies to support the T\&E infrastructure of fledgling companies, further driving innovation and progress in the field and especially in safety research. However, for countries that are already data-rich, such as the United States and China, incentives focused on data sharing may be insufficient. 

Third, it was unclear which incentives would encourage discussion and adoption of CBMs for different states. Disparities in military and technological capabilities among states may create resistance to CBMs, as some countries may believe that CBMs will disproportionately disadvantage them and data sharing incentives may be insufficient to overcome this perception. This belief may make a commitment to more intrusive monitoring and verification politically intractable in the near-term. There is an established literature in international relations that addresses the rational, normative, and psychological basis for participating in international agreements and arms control;\autocite{jervisArmsControlStability1993, brodieObjectivesArmsControl1976, bullArmsControlWorld1976, adlerEmergenceCooperationNational1992, wolfsthalWhyArmsControl2020, gottemoellerRethinkingNuclearArms2020, bettsSystemsPeaceCauses1992, larsenArmsControlCooperative2002, chayesInquiryWorkingsArms1972, adelmanArmsControlAgreements1984, millerPoliticsPromiseDomestic1984, daalderFutureArmsControl1992, krepsInstitutionalDesignArms2018, reddieGoverningInsecurityInstitutional2019, herzogNegotiationsUnderstandingMultilateral2021} this literature provides a foundation for the development of incentives that could target the adoption of CBMs for foundation models.

Despite the challenge of transnational coordination, confidence-building measures that target foundation models are important for international stability. Some of the suggestions in this document are already being developed as consumer protections, meaning that much of the remaining work will be on persuading parties—both private and public—that the adoption of CBMs will be beneficial for international security, Equally promising are the growing calls for international coordination on AI by governments, technology companies, and civil society in the face of state-to-state competitive tensions.\autocite{hicksOpinionWhatPentagon2023, clarkDODCommittedEthical2023, weiduoOpenAICEOCalls2023} These calls carve an opening for increased transnational dialogue between civil societies and scientific communities. As non-government actors become increasingly responsible for steering technologies that have global ramifications, many of the sociotechnical solutions that reduce misperception will need to be implemented at the technical layer and in collaboration with private actors.

\section*{Acknowledgements}
The authors would like to thank Wyatt Hoffman, Lauren Kahn, Philip Reiner, and Joel Parish for their valuable feedback on earlier versions of this manuscript.

\newpage
\phantomsection\addcontentsline{toc}{section}{References}
\printbibliography

@article{adelmanArmsControlAgreements1984,
  title = {Arms {{Control With}} and {{Without Agreements}}},
  author = {Adelman, Kenneth L.},
  year = {1984},
  month = dec,
  journal = {Foreign Affairs},
  number = {Winter 1984/85},
  issn = {0015-7120},
  urldate = {2023-07-18},
  langid = {american}
}

@article{adlerEmergenceCooperationNational1992,
  title = {The {{Emergence}} of {{Cooperation}}: {{National Epistemic Communities}} and the {{International Evolution}} of the {{Idea}} of {{Nuclear Arms Control}}},
  shorttitle = {The {{Emergence}} of {{Cooperation}}},
  author = {Adler, Emanuel},
  year = {1992},
  journal = {International Organization},
  volume = {46},
  number = {1},
  eprint = {2706953},
  eprinttype = {jstor},
  pages = {101--145},
  publisher = {{[MIT Press, University of Wisconsin Press, Cambridge University Press, International Organization Foundation]}},
  issn = {0020-8183},
  urldate = {2023-07-18}
}

@misc{AIAAIC,
  title = {{AIAAIC}},
  urldate = {2023-07-18},
  howpublished = {https://www.aiaaic.org/home},
  langid = {turkish}
}

@misc{AIIncidentsDatabase,
  title = {{{AI Incidents Database}}},
  journal = {Partnership on AI},
  urldate = {2023-07-18},
  howpublished = {https://partnershiponai.org/workstream/ai-incidents-database/},
  langid = {american}
}

@article{allynDeepfakeVideoZelenskyy2022,
  title = {Deepfake Video of {{Zelenskyy}} Could Be 'tip of the Iceberg' in Info War, Experts Warn},
  author = {Allyn, Bobby},
  year = {2022},
  month = mar,
  journal = {NPR},
  urldate = {2023-07-18},
  chapter = {Technology},
  langid = {english}
}

@misc{baiConstitutionalAIHarmlessness2022a,
  title = {Constitutional {{AI}}: {{Harmlessness}} from {{AI Feedback}}},
  shorttitle = {Constitutional {{AI}}},
  author = {Bai, Yuntao and Kadavath, Saurav and Kundu, Sandipan and Askell, Amanda and Kernion, Jackson and Jones, Andy and Chen, Anna and Goldie, Anna and Mirhoseini, Azalia and McKinnon, Cameron and Chen, Carol and Olsson, Catherine and Olah, Christopher and Hernandez, Danny and Drain, Dawn and Ganguli, Deep and Li, Dustin and {Tran-Johnson}, Eli and Perez, Ethan and Kerr, Jamie and Mueller, Jared and Ladish, Jeffrey and Landau, Joshua and Ndousse, Kamal and Lukosuite, Kamile and Lovitt, Liane and Sellitto, Michael and Elhage, Nelson and Schiefer, Nicholas and Mercado, Noemi and DasSarma, Nova and Lasenby, Robert and Larson, Robin and Ringer, Sam and Johnston, Scott and Kravec, Shauna and Showk, Sheer El and Fort, Stanislav and Lanham, Tamera and {Telleen-Lawton}, Timothy and Conerly, Tom and Henighan, Tom and Hume, Tristan and Bowman, Samuel R. and {Hatfield-Dodds}, Zac and Mann, Ben and Amodei, Dario and Joseph, Nicholas and McCandlish, Sam and Brown, Tom and Kaplan, Jared},
  year = {2022},
  month = dec,
  number = {arXiv:2212.08073},
  eprint = {2212.08073},
  primaryclass = {cs},
  publisher = {{arXiv}},
  doi = {10.48550/arXiv.2212.08073},
  urldate = {2023-07-18},
  archiveprefix = {arxiv}
}

@article{bettsSystemsPeaceCauses1992,
  title = {Systems for {{Peace}} or {{Causes}} of {{War}}? {{Collective Security}}, {{Arms Control}}, and the {{New Europe}}},
  shorttitle = {Systems for {{Peace}} or {{Causes}} of {{War}}?},
  author = {Betts, Richard K.},
  year = {1992},
  journal = {International Security},
  volume = {17},
  number = {1},
  eprint = {2539157},
  eprinttype = {jstor},
  pages = {5--43},
  publisher = {{The MIT Press}},
  issn = {0162-2889},
  doi = {10.2307/2539157},
  urldate = {2023-07-18}
}

@techreport{biden2022USNational2022,
  title = {2022 {{US National Security Strategy}}},
  author = {Biden, Joe},
  year = {2022},
  month = oct,
  address = {{Washington}},
  institution = {{The White House}},
  langid = {english}
}

@misc{billsLanguageModelsCan2023,
  title = {Language Models Can Explain Neurons in Language Models},
  author = {Bills, Steven and Cammarata, Nick and Mossing, Dan and Tillman, Henk and Gao, Leo and Wu, Jeff and Saunders, William},
  year = {2023},
  month = may,
  address = {{OpenAI}},
  collaborator = {Goh, Gabriel and Sutskever, Ilya and Leike, Jan}
}

@misc{bommasaniOpportunitiesRisksFoundation2022,
  title = {On the {{Opportunities}} and {{Risks}} of {{Foundation Models}}},
  author = {Bommasani, Rishi and Hudson, Drew A. and Adeli, Ehsan and Altman, Russ and Arora, Simran and {von Arx}, Sydney and Bernstein, Michael S. and Bohg, Jeannette and Bosselut, Antoine and Brunskill, Emma and Brynjolfsson, Erik and Buch, Shyamal and Card, Dallas and Castellon, Rodrigo and Chatterji, Niladri and Chen, Annie and Creel, Kathleen and Davis, Jared Quincy and Demszky, Dora and Donahue, Chris and Doumbouya, Moussa and Durmus, Esin and Ermon, Stefano and Etchemendy, John and Ethayarajh, Kawin and {Fei-Fei}, Li and Finn, Chelsea and Gale, Trevor and Gillespie, Lauren and Goel, Karan and Goodman, Noah and Grossman, Shelby and Guha, Neel and Hashimoto, Tatsunori and Henderson, Peter and Hewitt, John and Ho, Daniel E. and Hong, Jenny and Hsu, Kyle and Huang, Jing and Icard, Thomas and Jain, Saahil and Jurafsky, Dan and Kalluri, Pratyusha and Karamcheti, Siddharth and Keeling, Geoff and Khani, Fereshte and Khattab, Omar and Koh, Pang Wei and Krass, Mark and Krishna, Ranjay and Kuditipudi, Rohith and Kumar, Ananya and Ladhak, Faisal and Lee, Mina and Lee, Tony and Leskovec, Jure and Levent, Isabelle and Li, Xiang Lisa and Li, Xuechen and Ma, Tengyu and Malik, Ali and Manning, Christopher D. and Mirchandani, Suvir and Mitchell, Eric and Munyikwa, Zanele and Nair, Suraj and Narayan, Avanika and Narayanan, Deepak and Newman, Ben and Nie, Allen and Niebles, Juan Carlos and Nilforoshan, Hamed and Nyarko, Julian and Ogut, Giray and Orr, Laurel and Papadimitriou, Isabel and Park, Joon Sung and Piech, Chris and Portelance, Eva and Potts, Christopher and Raghunathan, Aditi and Reich, Rob and Ren, Hongyu and Rong, Frieda and Roohani, Yusuf and Ruiz, Camilo and Ryan, Jack and R{\'e}, Christopher and Sadigh, Dorsa and Sagawa, Shiori and Santhanam, Keshav and Shih, Andy and Srinivasan, Krishnan and Tamkin, Alex and Taori, Rohan and Thomas, Armin W. and Tram{\`e}r, Florian and Wang, Rose E. and Wang, William and Wu, Bohan and Wu, Jiajun and Wu, Yuhuai and Xie, Sang Michael and Yasunaga, Michihiro and You, Jiaxuan and Zaharia, Matei and Zhang, Michael and Zhang, Tianyi and Zhang, Xikun and Zhang, Yuhui and Zheng, Lucia and Zhou, Kaitlyn and Liang, Percy},
  year = {2022},
  month = jul,
  number = {arXiv:2108.07258},
  eprint = {2108.07258},
  primaryclass = {cs},
  publisher = {{arXiv}},
  doi = {10.48550/arXiv.2108.07258},
  urldate = {2023-07-17},
  archiveprefix = {arxiv}
}

@article{brodieObjectivesArmsControl1976,
  title = {On the {{Objectives}} of {{Arms Control}}},
  author = {Brodie, Bernard},
  year = {1976},
  journal = {International Security},
  volume = {1},
  number = {1},
  eprint = {2538574},
  eprinttype = {jstor},
  pages = {17--36},
  publisher = {{The MIT Press}},
  issn = {0162-2889},
  doi = {10.2307/2538574},
  urldate = {2023-07-18}
}

@misc{BuildingTrustHumanmachine,
  title = {Building Trust in Human-Machine Teams},
  journal = {Brookings},
  urldate = {2023-07-18},
  howpublished = {https://www.brookings.edu/articles/building-trust-in-human-machine-teams/},
  langid = {american}
}

@article{bullArmsControlWorld1976,
  title = {Arms {{Control}} and {{World Order}}},
  author = {Bull, Hedley},
  year = {1976},
  journal = {International Security},
  volume = {1},
  number = {1},
  eprint = {2538573},
  eprinttype = {jstor},
  pages = {3--16},
  publisher = {{The MIT Press}},
  issn = {0162-2889},
  doi = {10.2307/2538573},
  urldate = {2023-07-18}
}

@techreport{bureauofarmscontrolPoliticalDeclarationResponsible2023,
  title = {Political {{Declaration}} on {{Responsible Military Use}} of {{Artificial Intelligence}} and {{Autonomy}} - {{United States Department}} of {{State}}},
  author = {{Bureau of Arms Control}},
  year = {2023},
  month = feb,
  institution = {{U.S. Department of State}},
  urldate = {2023-07-17},
  langid = {american}
}

@misc{C2PASecurityConsiderations,
  title = {{{C2PA Security Considerations}} :: {{C2PA Specifications}}},
  journal = {C2PA},
  urldate = {2023-07-18},
  howpublished = {https://c2pa.org/specifications/specifications/1.0/security/Security\_Considerations.html\#\_provenance\_model}
}

@misc{CERTEUComputerEmergency,
  title = {{{CERT-EU}} \textendash{} Computer Emergency Response Team | {{European Union}}},
  urldate = {2023-07-18},
  howpublished = {https://european-union.europa.eu/institutions-law-budget/institutions-and-bodies/search-all-eu-institutions-and-bodies/computer-emergency-response-team-eu-institutions-bodies-and-agencies-cert-eu\_en},
  langid = {english}
}

@techreport{ChapterNSCAIFinal,
  title = {Chapter 4 - {{NSCAI Final Report}}},
  institution = {{National Security Commision on Artificial Intelligence}},
  urldate = {2023-07-17},
  langid = {english}
}

@article{chayesInquiryWorkingsArms1972,
  title = {An {{Inquiry}} into the {{Workings}} of {{Arms Control Agreements}}},
  author = {Chayes, Abram},
  year = {1972},
  journal = {Harvard Law Review},
  volume = {85},
  number = {5},
  eprint = {1339933},
  eprinttype = {jstor},
  pages = {905--969},
  publisher = {{The Harvard Law Review Association}},
  issn = {0017-811X},
  doi = {10.2307/1339933},
  urldate = {2023-07-18}
}

@misc{chrishmsftTransparencyNoteAzure2023,
  title = {Transparency {{Note}} for {{Azure OpenAI}} - {{Azure Cognitive Services}}},
  author = {ChrisHMSFT},
  year = {2023},
  month = may,
  urldate = {2023-07-18},
  howpublished = {https://learn.microsoft.com/en-us/legal/cognitive-services/openai/transparency-note},
  langid = {american}
}

@misc{clarkDODCommittedEthical2023,
  title = {{{DOD Committed}} to {{Ethical Use}} of {{Artificial Intelligence}}},
  author = {Clark, Joseph},
  year = {2023},
  month = jun,
  journal = {U.S. Department of Defense},
  urldate = {2023-07-18},
  howpublished = {https://www.defense.gov/News/News-Stories/Article/Article/3429864/dod-committed-to-ethical-use-of-artificial-intelligence/https\%3A\%2F\%2Fwww.defense.gov\%2FNews\%2FNews-Stories\%2FArticle\%2FArticle\%2F3429864\%2Fdod-committed-to-ethical-use-of-artificial-intelligence\%2F},
  langid = {american}
}

@techreport{CloseCallsNuclear2015,
  title = {Close {{Calls}} with {{Nuclear Weapons}}},
  year = {2015},
  month = jan,
  institution = {{Union of Concerned Scientists}},
  urldate = {2023-07-18},
  langid = {english}
}

@article{cohenAdvancedArtificialAgents2022,
  title = {Advanced Artificial Agents Intervene in the Provision of Reward},
  author = {Cohen, Michael K. and Hutter, Marcus and Osborne, Michael A.},
  year = {2022},
  journal = {AI Magazine},
  volume = {43},
  number = {3},
  pages = {282--293},
  issn = {2371-9621},
  doi = {10.1002/aaai.12064},
  urldate = {2023-07-18},
  copyright = {\textcopyright{} 2022 The Authors. AI Magazine published by Wiley Periodicals LLC on behalf of the Association for the Advancement of Artificial Intelligence.},
  langid = {english}
}

@misc{ConfidenceBuildingMeasuresCrossStrait,
  title = {Confidence-{{Building Measures}} | {{Cross-Strait Security Initiative}} | {{CSIS}}},
  urldate = {2023-07-17},
  howpublished = {https://www.csis.org/programs/international-security-program/isp-archives/asia-division/cross-strait-security-1},
  langid = {english}
}

@techreport{ConventionCertainConventional,
  title = {The {{Convention}} on {{Certain Conventional Weapons}} \textendash{} {{UNODA}}},
  institution = {{United Nations Office for Diasmament Affairs}},
  urldate = {2023-07-18},
  langid = {american}
}

@misc{CoreViewsAI,
  title = {Core {{Views}} on {{AI Safety}}: {{When}}, {{Why}}, {{What}}, and {{How}}},
  shorttitle = {Core {{Views}} on {{AI Safety}}},
  journal = {Anthropic},
  urldate = {2023-07-28},
  howpublished = {https://www.anthropic.com/index/core-views-on-ai-safety},
  langid = {english}
}

@misc{crawfordExcavatingAI,
  title = {Excavating {{AI}}},
  author = {Crawford, Kate and Paglen, Trevor},
  journal = {Excavating AI},
  urldate = {2023-07-19},
  howpublished = {https://excavating.ai},
  langid = {american}
}

@article{daalderFutureArmsControl1992,
  title = {The Future of Arms Control},
  author = {Daalder, Ivo H.},
  year = {1992},
  month = mar,
  journal = {Survival},
  volume = {34},
  number = {1},
  pages = {51--73},
  publisher = {{Routledge}},
  issn = {0039-6338},
  doi = {10.1080/00396339208442630},
  urldate = {2023-07-18}
}

@misc{DallE2SystemCard,
  title = {{{DallE-2 System Card}}},
  journal = {GitHub},
  urldate = {2023-07-18},
  howpublished = {https://github.com/openai/dalle-2-preview/blob/main/system-card.md},
  langid = {english}
}

@article{depasqualeCivilianPlanesShot2020,
  title = {Civilian {{Planes Shot Down}}: {{A Grim History}}},
  shorttitle = {Civilian {{Planes Shot Down}}},
  author = {DePasquale, Ron},
  year = {2020},
  month = jan,
  journal = {The New York Times},
  issn = {0362-4331},
  urldate = {2023-07-18},
  chapter = {World},
  langid = {american}
}

@article{dingRecentTrendsChina2023,
  title = {Recent {{Trends}} in {{China}}'s {{Large Language Model Landscape}}},
  author = {Ding, Jeffrey and Xiao, Jenny},
  year = {2023},
  month = apr,
  journal = {Centre for the Governance of AI},
  urldate = {2023-07-17},
  langid = {english}
}

@misc{DonovanAIpoweredDecisionmaking,
  title = {Donovan: {{AI-powered}} Decision-Making for Defense. | {{Scale AI}}},
  shorttitle = {Donovan},
  journal = {ScaleAI},
  urldate = {2023-07-17},
  howpublished = {https://scale.com/donovan}
}

@misc{duaDROPReadingComprehension2019,
  title = {{{DROP}}: {{A Reading Comprehension Benchmark Requiring Discrete Reasoning Over Paragraphs}}},
  shorttitle = {{{DROP}}},
  author = {Dua, Dheeru and Wang, Yizhong and Dasigi, Pradeep and Stanovsky, Gabriel and Singh, Sameer and Gardner, Matt},
  year = {2019},
  month = apr,
  number = {arXiv:1903.00161},
  eprint = {1903.00161},
  primaryclass = {cs},
  publisher = {{arXiv}},
  doi = {10.48550/arXiv.1903.00161},
  urldate = {2023-07-18},
  archiveprefix = {arxiv}
}

@article{duelferChronicMisperceptionInternational2011,
  title = {Chronic Misperception and International Conflict: {{The US-Iraq}} Experience.},
  author = {Duelfer, Charles A. and Dyson, Stephen Benedict},
  year = 2011,
  journal = {International Security},
  volume = {36},
  number = {4},
  pages = {73--100},
  issn = {1531-4804}
}

@misc{elhageToyModelsSuperposition2022,
  title = {Toy {{Models}} of {{Superposition}}},
  author = {Elhage, Nelson and Hume, Tristan and Olsson, Catherine and Schiefer, Nicholas and Henighan, Tom and Kravec, Shauna and {Hatfield-Dodds}, Zac and Lasenby, Robert and Drain, Dawn and Chen, Carol and Grosse, Roger and McCandlish, Sam and Kaplan, Jared and Amodei, Dario and Wattenberg, Martin and Olah, Christopher},
  year = {2022},
  month = sep,
  number = {arXiv:2209.10652},
  eprint = {2209.10652},
  primaryclass = {cs},
  publisher = {{arXiv}},
  doi = {10.48550/arXiv.2209.10652},
  urldate = {2023-07-18},
  archiveprefix = {arxiv}
}

@misc{englandAMPAuthenticationMedia2020,
  title = {{{AMP}}: {{Authentication}} of {{Media}} via {{Provenance}}},
  shorttitle = {{{AMP}}},
  author = {England, Paul and Malvar, Henrique S. and Horvitz, Eric and Stokes, Jack W. and Fournet, C{\'e}dric and {Burke-Aguero}, Rebecca and Chamayou, Amaury and Clebsch, Sylvan and Costa, Manuel and Deutscher, John and Erfani, Shabnam and Gaylor, Matt and Jenks, Andrew and Kane, Kevin and Redmiles, Elissa and Shamis, Alex and Sharma, Isha and Wenker, Sam and Zaman, Anika},
  year = {2020},
  month = jun,
  number = {arXiv:2001.07886},
  eprint = {2001.07886},
  primaryclass = {cs, eess},
  publisher = {{arXiv}},
  doi = {10.48550/arXiv.2001.07886},
  urldate = {2023-07-18},
  archiveprefix = {arxiv}
}

@techreport{EnsuringSafeSecure,
  title = {Ensuring {{Safe}}, {{Secure}}, and {{Trustworthy AI}}},
  address = {{Washington}},
  institution = {{The White House}}
}

@misc{Evals2023,
  title = {Evals},
  year = {2023},
  month = jul,
  urldate = {2023-07-18},
  copyright = {MIT},
  howpublished = {OpenAI}
}

@misc{ExploringPromptInjection2022,
  title = {Exploring {{Prompt Injection Attacks}}},
  year = {2022},
  month = dec,
  journal = {NCC Group Research Blog},
  urldate = {2023-07-18},
  howpublished = {https://research.nccgroup.com/2022/12/05/exploring-prompt-injection-attacks/},
  langid = {british}
}

@article{fearonRationalistExplanationsWar1995,
  title = {Rationalist {{Explanations}} for {{War}}},
  author = {Fearon, James D.},
  year = 1995,
  journal = {International Organization},
  volume = {49},
  number = {3},
  eprint = {2706903},
  eprinttype = {jstor},
  pages = {379--414}
}

@techreport{FormalInvestigationCircumstances1988,
  type = {Investigation {{Report}}},
  title = {Formal {{Investigation}} into the {{Circumstances Surrounding}} the {{Downing}} of {{Iran Air Flight}} 655 on 3 {{July}} 1988},
  year = {1988},
  month = jul,
  number = {93-FOI-0184},
  pages = {153},
  institution = {{U.S. Department of Defense}}
}

@misc{glaeseImprovingAlignmentDialogue2022a,
  title = {Improving Alignment of Dialogue Agents via Targeted Human Judgements},
  author = {Glaese, Amelia and McAleese, Nat and Tr{\k{e}}bacz, Maja and Aslanides, John and Firoiu, Vlad and Ewalds, Timo and Rauh, Maribeth and Weidinger, Laura and Chadwick, Martin and Thacker, Phoebe and {Campbell-Gillingham}, Lucy and Uesato, Jonathan and Huang, Po-Sen and Comanescu, Ramona and Yang, Fan and See, Abigail and Dathathri, Sumanth and Greig, Rory and Chen, Charlie and Fritz, Doug and Elias, Jaume Sanchez and Green, Richard and Mokr{\'a}, So{\v n}a and Fernando, Nicholas and Wu, Boxi and Foley, Rachel and Young, Susannah and Gabriel, Iason and Isaac, William and Mellor, John and Hassabis, Demis and Kavukcuoglu, Koray and Hendricks, Lisa Anne and Irving, Geoffrey},
  year = {2022},
  month = sep,
  number = {arXiv:2209.14375},
  eprint = {2209.14375},
  primaryclass = {cs},
  publisher = {{arXiv}},
  doi = {10.48550/arXiv.2209.14375},
  urldate = {2023-07-28},
  archiveprefix = {arxiv}
}

@misc{goldsteinGenerativeLanguageModels2023a,
  title = {Generative {{Language Models}} and {{Automated Influence Operations}}: {{Emerging Threats}} and {{Potential Mitigations}}},
  shorttitle = {Generative {{Language Models}} and {{Automated Influence Operations}}},
  author = {Goldstein, Josh A. and Sastry, Girish and Musser, Micah and DiResta, Renee and Gentzel, Matthew and Sedova, Katerina},
  year = {2023},
  month = jan,
  number = {arXiv:2301.04246},
  eprint = {2301.04246},
  primaryclass = {cs},
  publisher = {{arXiv}},
  doi = {10.48550/arXiv.2301.04246},
  urldate = {2023-07-18},
  archiveprefix = {arxiv}
}

@article{gottemoellerRethinkingNuclearArms2020,
  title = {Rethinking {{Nuclear Arms Control}}},
  author = {Gottemoeller, Rose},
  year = {2020},
  month = jul,
  journal = {The Washington Quarterly},
  volume = {43},
  number = {3},
  pages = {139--159},
  publisher = {{Routledge}},
  issn = {0163-660X},
  doi = {10.1080/0163660X.2020.1813382},
  urldate = {2023-07-18}
}

@misc{gowranAdobeLaunchesOpensource2022,
  title = {Adobe Launches Open-Source Tools to Tackle Visual Misinformation},
  author = {Gowran, Leigh Mc},
  year = {2022},
  month = jun,
  journal = {Silicon Republic},
  urldate = {2023-07-18},
  howpublished = {https://www.siliconrepublic.com/enterprise/adobe-digital-misinformation-cai-developer-tools},
  langid = {english}
}

@misc{GPT4SystemCard2023,
  title = {{{GPT-4 System Card}}},
  year = {2023},
  month = mar,
  publisher = {{OpenAI}},
  langid = {english}
}

@misc{H0201USSVincennes,
  title = {H-020-1: {{USS Vincennes Tragedy}}},
  shorttitle = {H-020-1},
  urldate = {2023-07-28},
  howpublished = {http://public1.nhhcaws.local/content/history/nhhc/about-us/leadership/director/directors-corner/h-grams/h-gram-020/h-020-1-uss-vincennes-tragedy--.html},
  langid = {american}
}

@misc{hendrycksMeasuringMassiveMultitask2023,
  title = {Measuring {{Massive Multitask Language Understanding}}},
  author = {Hendrycks, Dan},
  year = {2023},
  month = jul,
  urldate = {2023-07-18},
  copyright = {MIT}
}

@misc{hersmanWormholeEscalationNew2020,
  title = {Wormhole {{Escalation}} in the {{New Nuclear Age}}},
  author = {Hersman, Rebecca},
  year = {2020},
  month = jul,
  journal = {Texas National Security Review},
  urldate = {2023-07-18},
  howpublished = {https://tnsr.org/2020/07/wormhole-escalation-in-the-new-nuclear-age/},
  langid = {american}
}

@article{herzogNegotiationsUnderstandingMultilateral2021,
  title = {After the {{Negotiations}}: {{Understanding Multilateral Nuclear Arms Control}}},
  shorttitle = {After the {{Negotiations}}},
  author = {Herzog, Stephen},
  year = {2021},
  month = apr,
  journal = {Yale Graduate School of Arts and Sciences Dissertations}
}

@misc{hicksOpinionWhatPentagon2023,
  title = {Opinion | {{What}} the {{Pentagon Thinks About Artificial Intelligence}}},
  author = {Hicks, Kathleen},
  year = {2023},
  month = jun,
  journal = {POLITICO},
  urldate = {2023-07-18},
  howpublished = {https://www.politico.com/news/magazine/2023/06/15/pentagon-artificial-intelligence-china-00101751},
  langid = {english}
}

@misc{horowitzAIInternationalStability2021,
  title = {{{AI}} and {{International Stability}}: {{Risks}} and {{Confidence-Building Measures}}},
  shorttitle = {{{AI}} and {{International Stability}}},
  author = {Horowitz, Michael C. and Scharre, Paul},
  year = {2021},
  month = jan,
  urldate = {2023-07-18},
  howpublished = {https://www.cnas.org/publications/reports/ai-and-international-stability-risks-and-confidence-building-measures},
  langid = {english}
}

@article{horowitzFutureMilitaryApplications2020,
  title = {The {{Future}} of {{Military Applications}} of {{Artificial Intelligence}}: {{A Role}} for {{Confidence-Building Measures}}?},
  shorttitle = {The {{Future}} of {{Military Applications}} of {{Artificial Intelligence}}},
  author = {Horowitz, Michael C. and Kahn, Lauren and Mahoney, Casey},
  year = {2020},
  month = jan,
  journal = {Orbis},
  volume = {64},
  number = {4},
  pages = {528--543},
  issn = {0030-4387},
  doi = {10.1016/j.orbis.2020.08.003},
  urldate = {2023-07-18},
  langid = {english}
}

@article{horowitzLeadingArtificialIntelligence2021,
  title = {Leading in {{Artificial Intelligence}} through {{Confidence Building Measures}}},
  author = {Horowitz, Michael C. and Kahn, Lauren},
  year = {2021},
  month = oct,
  journal = {The Washington Quarterly},
  volume = {44},
  number = {4},
  pages = {91--106},
  publisher = {{Routledge}},
  issn = {0163-660X},
  doi = {10.1080/0163660X.2021.2018794},
  urldate = {2023-07-17}
}

@inproceedings{horvitzHorizonInteractiveCompositional2022a,
  title = {On the {{Horizon}}: {{Interactive}} and {{Compositional Deepfakes}}},
  shorttitle = {On the {{Horizon}}},
  booktitle = {{{INTERNATIONAL CONFERENCE ON MULTIMODAL INTERACTION}}},
  author = {Horvitz, Eric},
  year = {2022},
  month = nov,
  eprint = {2209.01714},
  primaryclass = {cs},
  pages = {653--661},
  doi = {10.1145/3536221.3558175},
  urldate = {2023-07-18},
  archiveprefix = {arxiv}
}

@misc{houseFACTSHEETBidenHarris2023,
  title = {{{FACT SHEET}}: {{Biden-Harris Administration Announces New Actions}} to {{Promote Responsible AI Innovation}} That {{Protects Americans}}' {{Rights}} and {{Safety}}},
  shorttitle = {{{FACT SHEET}}},
  author = {House, The White},
  year = {2023},
  month = may,
  journal = {The White House},
  urldate = {2023-07-18},
  howpublished = {https://www.whitehouse.gov/briefing-room/statements-releases/2023/05/04/fact-sheet-biden-harris-administration-announces-new-actions-to-promote-responsible-ai-innovation-that-protects-americans-rights-and-safety/},
  langid = {american}
}

@misc{Incident543Deepfake2020,
  title = {Incident 543: {{Deepfake}} of {{Explosion Near US Military Administration Building Reportedly Causes Stock Dip}}},
  shorttitle = {Incident 543},
  year = {2020},
  month = jan,
  urldate = {2023-07-18},
  howpublished = {https://incidentdatabase.ai/cite/543/},
  langid = {american}
}

@misc{InternationalTrafficArms2020,
  title = {International {{Traffic}} in {{Arms Regulations}}: {{U}}.{{S}}. {{Munitions List Categories I}}, {{II}}, and {{III}}},
  shorttitle = {International {{Traffic}} in {{Arms Regulations}}},
  year = {2020},
  month = jan,
  journal = {Federal Register},
  urldate = {2023-07-18},
  howpublished = {https://www.federalregister.gov/documents/2020/01/23/2020-00574/international-traffic-in-arms-regulations-us-munitions-list-categories-i-ii-and-iii}
}

@article{jervisArmsControlStability1993,
  title = {Arms {{Control}}, {{Stability}}, and {{Causes}} of {{War}}},
  author = {Jervis, Robert},
  year = {1993},
  journal = {Political Science Quarterly},
  volume = {108},
  number = {2},
  eprint = {2152010},
  eprinttype = {jstor},
  pages = {239--253},
  publisher = {{[Academy of Political Science, Wiley]}},
  issn = {0032-3195},
  doi = {10.2307/2152010},
  urldate = {2023-07-18}
}

@article{jervisCooperationSecurityDilemma1978,
  title = {Cooperation {{Under}} the {{Security Dilemma}}},
  author = {Jervis, Robert},
  year = {1978},
  month = jan,
  journal = {World Politics},
  volume = {30},
  number = {2},
  pages = {167--214},
  langid = {english}
}

@misc{jiangEvadingWatermarkBased2023,
  title = {Evading {{Watermark}} Based {{Detection}} of {{AI-Generated Content}}},
  author = {Jiang, Zhengyuan and Zhang, Jinghuai and Gong, Neil Zhenqiang},
  year = {2023},
  month = may,
  number = {arXiv:2305.03807},
  eprint = {2305.03807},
  primaryclass = {cs},
  publisher = {{arXiv}},
  doi = {10.48550/arXiv.2305.03807},
  urldate = {2023-07-18},
  archiveprefix = {arxiv}
}

@article{johnsonArtificialIntelligenceDrone2020a,
  title = {Artificial {{Intelligence}}, {{Drone Swarming}} and {{Escalation Risks}} in {{Future Warfare}}},
  author = {Johnson, James},
  year = {2020},
  month = feb,
  journal = {The RUSI Journal},
  volume = {165},
  number = {2},
  pages = {26--36},
  publisher = {{Routledge}},
  issn = {0307-1847},
  doi = {10.1080/03071847.2020.1752026},
  urldate = {2023-07-18}
}

@article{knightAutonomousWeaponsAre,
  title = {Autonomous {{Weapons Are Here}}, but the {{World Isn}}'t {{Ready}} for {{Them}}},
  author = {Knight, Will},
  journal = {Wired},
  issn = {1059-1028},
  urldate = {2023-07-18},
  chapter = {tags},
  langid = {american}
}

@article{krepsInstitutionalDesignArms2018,
  title = {The {{Institutional Design}} of {{Arms Control Agreements}}},
  author = {Kreps, Sarah E.},
  year = {2018},
  month = jan,
  journal = {Foreign Policy Analysis},
  volume = {14},
  number = {1},
  pages = {127--147},
  issn = {1743-8586},
  doi = {10.1093/fpa/orw045},
  urldate = {2023-07-18}
}

@article{lafranchiUSChinaConundrumCan2023,
  title = {{{US-China}} Conundrum: {{Can}} Hotline Diplomacy Work If Trust Isn't a Goal?},
  shorttitle = {{{US-China}} Conundrum},
  author = {LaFranchi, Howard},
  year = {2023},
  month = mar,
  journal = {Christian Science Monitor},
  issn = {0882-7729},
  urldate = {2023-07-18}
}

@book{larsenArmsControlCooperative2002,
  title = {Arms Control: Cooperative Security in a Changing Environment},
  shorttitle = {Arms Control},
  editor = {Larsen, Jeffrey Arthur},
  year = {2002},
  publisher = {{Lynne Rienner Publishers}},
  address = {{Boulder, Colo}},
  isbn = {978-1-58826-013-0},
  lccn = {JZ5687 .A764 2002}
}

@misc{leikeOurApproachAlignment2022,
  title = {Our Approach to Alignment Research},
  author = {Leike, Jan and Schulman, John and Wu, Jeffrey},
  year = {2022},
  month = aug,
  urldate = {2023-07-28},
  howpublished = {https://openai.com/blog/our-approach-to-alignment-research},
  langid = {american}
}

@misc{LockedShields,
  title = {Locked {{Shields}}},
  journal = {CCDCOE},
  urldate = {2023-07-18},
  howpublished = {https://ccdcoe.org/exercises/locked-shields/}
}

@misc{lohnEstimatingBrittlenessAI2020,
  title = {Estimating the {{Brittleness}} of {{AI}}: {{Safety Integrity Levels}} and the {{Need}} for {{Testing Out-Of-Distribution Performance}}},
  shorttitle = {Estimating the {{Brittleness}} of {{AI}}},
  author = {Lohn, Andrew J.},
  year = {2020},
  month = sep,
  number = {arXiv:2009.00802},
  eprint = {2009.00802},
  primaryclass = {cs, stat},
  publisher = {{arXiv}},
  doi = {10.48550/arXiv.2009.00802},
  urldate = {2023-07-18},
  archiveprefix = {arxiv}
}

@misc{loweAligningLanguageModels2022,
  title = {Aligning Language Models to Follow Instructions},
  author = {Lowe, Ryan and Leike, Jan},
  year = {2022},
  month = jan,
  urldate = {2023-07-18},
  howpublished = {https://openai.com/research/instruction-following},
  langid = {american}
}

@misc{MicrosoftTuringAcademic,
  title = {Microsoft {{Turing Academic Program}} ({{MS-TAP}})},
  journal = {Microsoft Research},
  urldate = {2023-07-18},
  langid = {american}
}

@article{millerPoliticsPromiseDomestic1984,
  title = {Politics over {{Promise}}: {{Domestic Impediments}} to {{Arms Control}}},
  shorttitle = {Politics over {{Promise}}},
  author = {Miller, Steven E.},
  year = {1984},
  journal = {International Security},
  volume = {8},
  number = {4},
  eprint = {2538563},
  eprinttype = {jstor},
  pages = {67--90},
  publisher = {{The MIT Press}},
  issn = {0162-2889},
  doi = {10.2307/2538563},
  urldate = {2023-07-18}
}

@article{milmoUKInvest900m2023,
  title = {{{UK}} to Invest \textsterling 900m in Supercomputer in Bid to Build Own `{{BritGPT}}'},
  author = {Milmo, Dan and Hern, Alex},
  year = {2023},
  month = mar,
  journal = {The Guardian},
  issn = {0261-3077},
  urldate = {2023-07-17},
  chapter = {Technology},
  langid = {british}
}

@inproceedings{mitchellModelCardsModel2019a,
  title = {Model {{Cards}} for {{Model Reporting}}},
  booktitle = {Proceedings of the {{Conference}} on {{Fairness}}, {{Accountability}}, and {{Transparency}}},
  author = {Mitchell, Margaret and Wu, Simone and Zaldivar, Andrew and Barnes, Parker and Vasserman, Lucy and Hutchinson, Ben and Spitzer, Elena and Raji, Inioluwa Deborah and Gebru, Timnit},
  year = {2019},
  month = jan,
  eprint = {1810.03993},
  primaryclass = {cs},
  pages = {220--229},
  doi = {10.1145/3287560.3287596},
  urldate = {2023-07-19},
  archiveprefix = {arxiv}
}

@techreport{morganMilitaryApplicationsArtificial2020,
  title = {Military {{Applications}} of {{Artificial Intelligence}}: {{Ethical Concerns}} in an {{Uncertain World}}},
  shorttitle = {Military {{Applications}} of {{Artificial Intelligence}}},
  author = {Morgan, Forrest E. and Boudreaux, Benjamin and Lohn, Andrew J. and Ashby, Mark and Curriden, Christian and Klima, Kelly and Grossman, Derek},
  year = {2020},
  month = apr,
  institution = {{RAND Corporation}},
  urldate = {2023-07-18},
  langid = {english}
}

@misc{NTSBAccidentReport2019,
  title = {{{NTSB Accident Report}} on {{Fatal}} 2017 {{USS John McCain Collision}} off {{Singapore}}},
  year = {2019},
  month = aug,
  journal = {USNI News},
  urldate = {2023-07-18},
  chapter = {Documents},
  langid = {american}
}

@article{oltermannEuropeanPoliticiansDuped2022,
  title = {European Politicians Duped into Deepfake Video Calls with Mayor of {{Kyiv}}},
  author = {Oltermann, Philip},
  year = {2022},
  month = jun,
  journal = {The Guardian},
  issn = {0261-3077},
  urldate = {2023-07-18},
  chapter = {World news},
  langid = {british}
}

@techreport{OpinionIntegrationAutonomy2021,
  title = {Opinion {{On The Integration Of Autonomy Into Lethal Weapon Systems}}},
  year = {2021},
  month = apr,
  pages = {5--6},
  institution = {{Minist\'ere Des Arm\'ees Defence Ethics Committee}}
}

@misc{OverviewC2PA,
  title = {Overview - {{C2PA}}},
  journal = {C2PA},
  urldate = {2023-07-18},
  howpublished = {https://c2pa.org/}
}

@misc{PAIResponsiblePractices,
  title = {{{PAI}}'s {{Responsible Practices}} for {{Synthetic Media}}},
  journal = {Partnership on AI - Synthetic Media},
  urldate = {2023-07-28},
  howpublished = {https://syntheticmedia.partnershiponai.org/},
  langid = {american}
}

@misc{PalantirArtificialIntelligence,
  title = {Palantir {{Artificial Intelligence Platform}}},
  journal = {Palantir},
  urldate = {2023-07-17},
  howpublished = {https://www.palantir.com/platforms/aip/},
  langid = {english}
}

@misc{PapersCodeMMLU,
  title = {Papers with {{Code}} - {{MMLU Benchmark}} ({{Multi-task Language Understanding}})},
  urldate = {2023-07-18},
  howpublished = {https://paperswithcode.com/sota/multi-task-language-understanding-on-mmlu},
  langid = {english}
}

@article{puscasConfidenceBuildingMeasuresArtificial2022,
  title = {Confidence-{{Building Measures}} for {{Artificial Intelligence}}: {{A Framing Paper}}},
  shorttitle = {Confidence-{{Building Measures}} for {{Artificial Intelligence}}},
  author = {Puscas, Iona},
  year = {2022},
  journal = {United Nations Institute for Disarmament Research},
  urldate = {2023-07-17},
  langid = {english}
}

@inproceedings{rajiClosingAIAccountability2020,
  title = {Closing the {{AI}} Accountability Gap: Defining an End-to-End Framework for Internal Algorithmic Auditing},
  shorttitle = {Closing the {{AI}} Accountability Gap},
  booktitle = {Proceedings of the 2020 {{Conference}} on {{Fairness}}, {{Accountability}}, and {{Transparency}}},
  author = {Raji, Inioluwa Deborah and Smart, Andrew and White, Rebecca N. and Mitchell, Margaret and Gebru, Timnit and Hutchinson, Ben and {Smith-Loud}, Jamila and Theron, Daniel and Barnes, Parker},
  year = {2020},
  month = jan,
  series = {{{FAT}}* '20},
  pages = {33--44},
  publisher = {{Association for Computing Machinery}},
  address = {{New York, NY, USA}},
  doi = {10.1145/3351095.3372873},
  urldate = {2023-07-18},
  isbn = {978-1-4503-6936-7}
}

@phdthesis{reddieGoverningInsecurityInstitutional2019,
  title = {Governing {{Insecurity}}: {{Institutional Design}}, {{Compliance}}, and {{Arms Control}}},
  shorttitle = {Governing {{Insecurity}}},
  author = {Reddie, Andrew William},
  year = {2019},
  urldate = {2023-07-18},
  langid = {english},
  school = {UC Berkeley}
}

@misc{RepositoryMilitaryConfidencebuilding,
  title = {Repository of Military Confidence-Building Measures \textendash{} {{UNODA}}},
  urldate = {2023-07-18},
  langid = {american}
}

@article{ridAttributingCyberAttacks2015,
  title = {Attributing {{Cyber Attacks}}},
  author = {Rid, Thomas and Buchanan, Ben},
  year = {2015},
  journal = {The Journal of Strategic Studies},
  volume = {38},
  number = {1-2},
  pages = {4--37},
  doi = {10.1080/01402390.2014.977382}
}

@misc{sadasivanCanAIGeneratedText2023,
  title = {Can {{AI-Generated Text}} Be {{Reliably Detected}}?},
  author = {Sadasivan, Vinu Sankar and Kumar, Aounon and Balasubramanian, Sriram and Wang, Wenxiao and Feizi, Soheil},
  year = {2023},
  month = jun,
  number = {arXiv:2303.11156},
  eprint = {2303.11156},
  primaryclass = {cs},
  publisher = {{arXiv}},
  doi = {10.48550/arXiv.2303.11156},
  urldate = {2023-07-18},
  archiveprefix = {arxiv}
}

@misc{shahGoalMisgeneralizationWhy2022,
  title = {Goal {{Misgeneralization}}: {{Why Correct Specifications Aren}}'t {{Enough For Correct Goals}}},
  shorttitle = {Goal {{Misgeneralization}}},
  author = {Shah, Rohin and Varma, Vikrant and Kumar, Ramana and Phuong, Mary and Krakovna, Victoria and Uesato, Jonathan and Kenton, Zac},
  year = {2022},
  month = nov,
  number = {arXiv:2210.01790},
  eprint = {2210.01790},
  primaryclass = {cs},
  publisher = {{arXiv}},
  doi = {10.48550/arXiv.2210.01790},
  urldate = {2023-07-18},
  archiveprefix = {arxiv}
}

@book{snyderDeterrenceDefense1961,
  title = {Deterrence and {{Defense}}},
  author = {Snyder, Glenn Herald},
  year = {1961},
  publisher = {{Princeton University Press}},
  urldate = {2023-07-18},
  isbn = {978-0-691-65209-2},
  langid = {english}
}

@techreport{sunakIntegratedReviewRefresh2023,
  type = {{{UK HM Government Report}}},
  title = {Integrated {{Review Refresh}} 2023},
  author = {Sunak, Rishi},
  year = {2023},
  month = mar,
  pages = {56},
  institution = {{HM Government}},
  isbn = {978-1-5286-3962-0},
  langid = {british}
}

@misc{tamGentleIntroductionHallucinations2023,
  title = {A {{Gentle Introduction}} to {{Hallucinations}} in {{Large Language Models}}},
  author = {Tam, Adrian},
  year = {2023},
  month = jun,
  journal = {MachineLearningMastery.com},
  urldate = {2023-07-18},
  langid = {american}
}

@misc{tonerWhatAreGenerative2023,
  title = {What {{Are Generative AI}}, {{Large Language Models}}, and {{Foundation Models}}?},
  author = {Toner, Helen},
  year = {2023},
  month = may,
  journal = {Center for Security and Emerging Technology},
  urldate = {2023-07-18},
  langid = {american}
}

@unpublished{TreatyOpenSkies1992,
  type = {Treaty},
  title = {Treaty on {{Open Skies}}},
  year = {1992},
  month = mar,
  address = {{Helsinki}}
}

@misc{USCERTUnitedStates,
  title = {{{US-CERT}} ({{United States Computer Emergency Readiness Team}}) - {{Glossary}} | {{CSRC}}},
  urldate = {2023-07-18},
  howpublished = {https://csrc.nist.gov/glossary/term/us\_cert},
  langid = {american}
}

@misc{wardShaheenAdminGet2023,
  title = {Shaheen to Admin: {{Get}} Me the {{Black Sea}} Strategy},
  shorttitle = {Shaheen to Admin},
  author = {Ward, Alexander and Berg, Matt and Ukenye, Lawrence},
  year = {2023},
  month = jul,
  journal = {POLITICO},
  urldate = {2023-07-17},
  howpublished = {https://www.politico.com/newsletters/national-security-daily/2023/03/21/shaheen-to-admin-get-me-the-black-sea-strategy-00088048},
  langid = {english}
}

@misc{wehsenerAINC3IntegrationAdversarial,
  title = {{{AI-NC3 Integration}} in an {{Adversarial Context}}: {{Strategic Stability Risks}} and {{Confidence Building Measures}}},
  shorttitle = {{{AI-NC3 Integration}} in an {{Adversarial Context}}},
  author = {Wehsener, Alexa and Reddie, Andrew W. and Walker, Leah and Reiner, Philip},
  journal = {Institute for Security and Technology},
  urldate = {2023-07-18},
  howpublished = {https://securityandtechnology.org/virtual-library/reports/ai-nc3-integration-in-an-adversarial-context-strategic-stability-risks-and-confidence-building-measures/}
}

@misc{weiduoOpenAICEOCalls2023,
  title = {{{OpenAI CEO}} Calls for Global Cooperation on {{AI}} Regulation, Says `{{China}} Has Some of the Best {{AI}} Talent in the World' - {{Global Times}}},
  author = {Weiduo, Shen},
  year = {2023},
  month = jan,
  journal = {Global Times},
  urldate = {2023-07-18},
  howpublished = {https://www.globaltimes.cn/page/202306/1292326.shtml}
}

@misc{whitakerWhenRussianHackers2018,
  title = {When {{Russian}} Hackers Targeted the {{U}}.{{S}}. Election Infrastructure},
  author = {Whitaker, Bill},
  year = {2018},
  month = jul,
  journal = {CBS News},
  urldate = {2023-07-18},
  howpublished = {https://www.cbsnews.com/news/when-russian-hackers-targeted-the-u-s-election-infrastructure/},
  langid = {american}
}

@misc{wolfBackingWhyRed2016,
  title = {Backing {{Down}}: {{Why Red Lines Matter}} in {{Geopolitics}}},
  shorttitle = {Backing {{Down}}},
  author = {Wolf, Albert},
  year = {2016},
  month = aug,
  journal = {Modern War Institute},
  urldate = {2023-07-18},
  howpublished = {https://mwi.westpoint.edu/geopolitical-costs-red-lines/},
  langid = {american}
}

@article{wolfsthalWhyArmsControl2020,
  title = {Why {{Arms Control}}?},
  author = {Wolfsthal, Jon Brook},
  year = {2020},
  journal = {Daedalus},
  volume = {149},
  number = {2},
  eprint = {48591315},
  eprinttype = {jstor},
  pages = {101--115},
  publisher = {{[American Academy of Arts \& Sciences, The MIT Press]}},
  issn = {0011-5266},
  urldate = {2023-07-18}
}

@misc{wongWhatAmericansAre2023,
  title = {What {{Americans}} Are Really Excited about \textemdash{} and Scared of \textemdash{} When It Comes to {{AI}}},
  author = {Wong, Edwin H.},
  year = {2023},
  month = jun,
  journal = {Vox Media},
  urldate = {2023-07-28},
  howpublished = {https://www.voxmedia.com/2023/6/26/23769834/what-americans-are-really-excited-about-and-scared-of-when-it-comes-to-ai},
  langid = {english}
}

@misc{wuRecursivelySummarizingBooks2021,
  title = {Recursively {{Summarizing Books}} with {{Human Feedback}}},
  author = {Wu, Jeff and Ouyang, Long and Ziegler, Daniel M. and Stiennon, Nisan and Lowe, Ryan and Leike, Jan and Christiano, Paul},
  year = {2021},
  month = sep,
  number = {arXiv:2109.10862},
  eprint = {2109.10862},
  primaryclass = {cs},
  publisher = {{arXiv}},
  doi = {10.48550/arXiv.2109.10862},
  urldate = {2023-07-18},
  archiveprefix = {arxiv}
}

@misc{zhouLeasttoMostPromptingEnables2023,
  title = {Least-to-{{Most Prompting Enables Complex Reasoning}} in {{Large Language Models}}},
  author = {Zhou, Denny and Sch{\"a}rli, Nathanael and Hou, Le and Wei, Jason and Scales, Nathan and Wang, Xuezhi and Schuurmans, Dale and Cui, Claire and Bousquet, Olivier and Le, Quoc and Chi, Ed},
  year = {2023},
  month = apr,
  number = {arXiv:2205.10625},
  eprint = {2205.10625},
  primaryclass = {cs},
  publisher = {{arXiv}},
  doi = {10.48550/arXiv.2205.10625},
  urldate = {2023-07-18},
  archiveprefix = {arxiv}
}

@article{zieglerAdversarialTrainingHighstakes2022,
  title = {Adversarial Training for High-Stakes Reliability},
  author = {Ziegler, Daniel and Nix, Seraphina and Chan, Lawrence and Bauman, Tim and {Schmidt-Nielsen}, Peter and Lin, Tao and Scherlis, Adam and Nabeshima, Noa and {Weinstein-Raun}, Benjamin and {de Haas}, Daniel and Shlegeris, Buck and Thomas, Nate},
  year = {2022},
  month = dec,
  journal = {Advances in Neural Information Processing Systems},
  volume = {35},
  pages = {9274--9286},
  urldate = {2023-07-18},
  langid = {english}
}

\end{document}